# Rejoinder: Microarrays, Empirical Bayes and the Two-Groups Model

**Bradley Efron**

The Fisher–Neyman–Pearson theory of hypothesis testing was a triumph of mathematical elegance and practical utility. It was never designed, though, to handle 10,000 tests at once, and one can see contemporary statisticians struggling to develop theories appropriate to our new scientific environment. This paper is part of that effort: starting from just the two-groups model (2.1), it aims to show Bayesian and frequentist ideas merging into a practical framework for large-scale simultaneous testing.

False discovery rates, Benjamini and Hochberg's influential contribution to modern statistical theory, is the main methodology featured in the paper, but I really was not trying to sell any specific technology as the final word. In fact, the discussants offer an attractive menu of alternatives. It is still early in the large-scale hypothesis testing story, and I expect, and hope for, major developments in both theory and practice.

The central issue, as Carl Morris makes clear, is the combination of information from a collection of more or less similar sources, for example from the expression levels of different genes in a microarray study. Crucial questions revolve around the comparability and relevance of the various sources, as well as the proper choice of a null distribution. Technical issues such as the exact control of Type I errors are important as well, but, in my opinion, have played too big a role in the microarray literature. The discussions today are an appealing mixture of technical facility and big-picture thinking. They are substantial essays in their own right, and I will be able to respond here to only a few of the issues raised.

*Bradley Efron is Professor, Department of Statistics, Stanford University, Stanford, California 94305, USA e-mail: brad@stat.stanford.edu.*

I once wrote, about the jackknife, that *good* simple ideas are our most precious intellectual commodity. False discovery rates fall into that elite category. The two-groups model is used here to unearth the Bayesian roots of Benjamini and Hochberg's originally frequentist construction. In a Bayesian framework it is natural to focus on local false discovery rates, $\text{fdr}(z)$, rather than the original tail area version $\text{Fdr}(z)$. My apologies to Professor Benjamini for seeming to suggest that fdr is more immune than Fdr to correlations between the $z$-values. All false discovery rates are basically ratios of expectations, and as such remain relatively unbiased in the face of correlation. It is only the proof of the exact Fdr control property that involves some form of independence.

In the same spirit, I have to disagree that Fdr produces more reproducible results than fdr. Both methods operate at the mercy of an experiment's power, and low-power situations, such as the prostate cancer study, are certain to produce highly variable lists of "significant" cases. (At this point, let me repeat my plea for a better term than "significant" for the cases found to be nonnull, a dubious nomenclature even in classical settings, and definitely misleading for large-scale testing.)

As suggested by Figure 2, there is no great conceptual difference between fdr and Fdr, nor have I found much difference in applications. Table 1 says something about their comparative estimation accuracy. As Professor Cai suggests, the statistician can combine the two, using Fdr to select a reportable list of nonnull candidates, and fdr to differentiate the level of certainty within the list. Here the two roles reflect Benjamini's distinction between decision theory and inference, that is, between making a firm choice of nonnull cases and providing an estimate of just how nonnull they are.

As an enthusiastic collector of reasons to distrust the theoretical null distribution, I am happy to add *preselection of cases* to the list. Professor Benjamini correctly points out the dangers of this practice— among other things, it deprives the statistician of crucial evidence about the null distribution. If questioning the theoretical null seems heretical, it is worth remembering similar questions arising in classical ANOVA applications, for instance whether to use $\sigma^2$ (error) or $\sigma^2$ (interaction) in assessing the main effects of a two-way table. I share Benjamini's preference for finding the "right" theoretical null, but that is the counsel of perfection, often unattainable in examples like the education data.

Questions of exchangeability play a key role in large-scale hypothesis testing, as emphasized in Professor Morris's nice essay. The answer to "Which problems should be tested together?" is not always "All the ones the investigator put on my desk." A paper written after this article, "Simultaneous inference: When should hypothesis testing problems be combined?" (Efron, 2008) attacks this problem without conquering it. As Morris points out, covariates like school size in the education example may undercut exchangeability—the nonnull $z$-values for larger schools might lie farther away from 0. My paper suggests how to incorporate covariates into an efficient fdr analysis.

In this paper, only the paragraph following that of (3.10) has anything to say about exchangeability. (Notice that the local fdr puts less strain on exchangeability than tail-area Fdr since only the cases near some particular value $z$ are considered together.) For the education example we might be willing to accept exchangeability for the null $z_i$'s, from simple binomial calculations, though not for the nonnull cases. The interpretation of the equivalent of "2.68/17" in the paragraph following (3.10) could thus be modified in a Bayesian way to assign greater nonnull probability to the larger schools.

Morris' Section 3 is especially pertinent. His formula for $p(\mu|z)$ is related to my discussion of the Benjamini–Yekutieli False Coverage Rate method in Section 7, particularly (7.2)–(7.4). Originally I had hoped to develop an empirical Bayes method for estimating such models, but the effort foundered on practical difficulties involving the perils of deconvolution.

Section 6 on the "one-group model" is the ugly duckling of the current paper, but it bears on some important points raised in the discussion. Figure 7 concerns a fuzzy version of simultaneous hypothesis testing, where, as in Morris' hospital example, the usual single-point null hypotheses seem unequal to the task. The development from (6.6) onwards, particularly (6.12), bears on the possibility of nonnormal null distributions, and is about as far as I can go in answering Professors Rice and Spiegelhalter's penultimate question.

With $g(\mu)$ a normal distribution, model (6.1) returns us to the realm of Stein estimation, the scene of my happy collaborations with Carl Morris. I continue to be surprised at how much harder bumpy, nonnormal models like (7.1) are to deal with. James–Stein estimation works fine with, say, $N = 10$ component problems, while the Robbins' form of empirical Bayes appropriate to (7.1) seems to require hundreds or thousands. The information calculations in Efron (2008) reinforce this gloomy assessment. Maybe I am trying to be overly nonparametric in constructing the empirical Bayes Fdr estimates, but it is hard to imagine a generally satisfactory parametric formulation for (6.1). Or perhaps it is just that hypothesis testing is more demanding than estimation.

Rice and Spiegelhalter propose an attractive algorithm: rather than modeling the marginal density $f(z)$ as in (3.6), they suggest directly modeling fdr$(z)$. The resulting Huber form for $f(z)$ has a pleasant appearance, and I was relieved to see their results agreeing with mine.

The Rice–Spiegelhalter model involves only two free parameters, $k_a$ and $k_b$, as opposed to seven in (3.6). I doubt that two will be enough to cover a general range of applications, but would be happy to be proved wrong. For example, it might sometimes be necessary to have different exponential rates of decay in the two tails, rather than forcing them to be the same. [Perhaps I am just trying to lob the "ad hoc" accusation back into Rice and Spiegelhalter's court. Equations (3.4)–(3.6) describe a standard Poisson regression model; users of *locfdr* can select the degree of the regression, seven being only the default.] In any case, the direct modeling of fdr$(z)$ is a promising new route of attack.

"Efficiency" in Professor Cai's essay is what I called "power" in Section 3, a somewhat neglected aspect of multiple testing that now seems to be attracting attention. My diagnostic $E\widehat{\text{fdr}}^{(1)}$, (3.9), is trying to estimate the power parameter

$$1 - \int \text{fnr}(z) \cdot f_1(z)\, dz,$$

where fnr$(z)$ is the "local false nondiscovery rate" $1 - \text{fdr}(z)$, to use Cai's terminology. See Efron (2007).

Usually $\widehat{\text{fdr}}(z)$ declines monotonically as we move away from $z = 0$ in either direction, so that in each tail $\widehat{\text{fdr}}(z_i)$ orders evidence against the null in the same way as the $p$-value, $p_i$. The ordering can be different, however, if we try to compare evidence across the two tails. Cai's results, with Sun, show that it is better to define the decision boundary in terms of fdr-values than $p$-values, for example by $\widehat{\mathrm{fdr}}(z_i) \leq 0.2$ rather than using a $p$-value cutoff. This nicely reinforces the utility of the Bayesian quantity $\mathrm{fdr}(z)$ (2.7) for frequentist decision-theoretic calculations.

Jin and Cai have a quite different method for empirical null estimation, based on Fourier analysis. This moves in the opposite direction from Rice and Spiegelhalter, more nonparametric rather than less, and again seems to give good estimates.

Large-scale statistical inference blurs the line between Bayesians and frequentists: Bayesian information accumulates, and cannot be ignored, but the accumulation itself favors the use of frequentist tactics. The definition of "empirical Bayes," if there is one, lies somewhere in the realm of Bayesian–frequentist cooperation. Morris points out that this broad-sense definition of empirical Bayes was too wide for Robbins, and maybe for him too, but it is probably enough for the methodological goals of this paper.

My thanks go to the discussants, and also to the editor Ed George for organizing a session on this lively topic.